\documentclass[10pt,aps,prl,nofootinbib,superscriptaddress,twocolumn,preprintnumbers,balancelastpage]{revtex4}
\usepackage{amsmath}
\usepackage{graphicx}
\usepackage{caption}
\usepackage{subcaption}
\begin{document}
\title{High Resolution Muon Computed Tomography at Neutrino Beam Facilities}
\author{Burkhant Suerfu}
\email{suerfu@princeton.edu}
\affiliation{Department of Physics, Princeton University, Princeton, NJ 08544}
\author{Christopher G. Tully}
\affiliation{Department of Physics, Princeton University, Princeton, NJ 08544}
\date{\today}

\begin{abstract}
X-ray computed tomography (CT) has an indispensable role in constructing 3D images of objects
made from light materials. However, limited by absorption coefficients, X-rays cannot deeply penetrate materials such as copper and lead. Here we show via simulation that muon beams can provide high resolution tomographic images of dense objects and of structures within the interior of dense objects. The effects of resolution broadening from multiple scattering diminish with increasing muon momentum.  As the momentum of the muon increases, the contrast of the image goes down and therefore requires higher resolution in the muon spectrometer to resolve the image.  The variance of the measured muon momentum reaches a minimum and then increases with increasing muon momentum.  The impact of the increase in variance is to require a higher integrated muon flux to reduce fluctuations.
The flux requirements and level of contrast needed for high resolution muon computed tomography are well matched to the muons produced in the pion decay pipe at a neutrino beam facility and what can be achieved for momentum resolution in a muon spectrometer.
Such an imaging system can be applied in archaeology, art history, engineering, material identification and whenever there is a need to image inside a transportable object constructed of dense materials. 
\end{abstract}
\maketitle

\section{Introduction}
Computed tomographic (CT) images are formed by combining X-ray projection images from multiple angles, each of which record the amount of X-ray absorption as function of position, which is proportional to line-integrated X-ray absorption coefficient\cite{ct-principle}. However, the application of CT is limited by the large absorption coefficient of X-rays in certain materials such as heavy metals. For example, for $100$ keV X-rays the attenuation coefficient is $\approx 3.5$ cm$^{-1}$\cite{nist}. After $2$~cm of copper plate, the intensity is reduced to $0.1$\%.  If the object is significantly thicker, then the required power of the X-ray generator increases exponentially\cite{ct-in-archaeology}.\\
The idea of using muons for imaging dense objects dates back to Luis Alvarez and collaborators, who used cosmic muons to search for hidden chambers in the ancient Egyptian pyramids\cite{alvarez-pyramid}.
Cosmic muons continue to be used for diverse applications from geological information from imaging volcanos\cite{volcano-japan, mu-ray} to commercial and security use in the detection of dense radioactive materials in cargo containers\cite{multiple-scattering-high-z,material-id, industrial-structure}.  Most cosmic muon imaging experiments are, however, counting experiments and the muon energy and direction are not well controlled.  The cosmic muon flux is not suitable for imaging small objects with high resolution.  Therefore, we would like to investigate the possibilities and limitations of using muon beams from an accelerator complex for the purpose of high resolution tomographic imaging.\\

Muons incident on matter with electron number density $n$, atomic number $z$, and mean ionization energy $I$ will lose energy according to Bethe-Bloch equation\cite{bethe-bloch}:
	\begin{equation}
		-\frac{dE}{dx}=\frac{4\pi n z^2 e^4}{m_ev^2}[\log(\frac{2m_ev^2\gamma^2}{I})-\beta^2].\label{bethe-block}
	\end{equation}
Depending on the internal structure of the object, muons incident on different regions will traverse materials of different types and densities, and as a result come out with different mean energies.  If the mean energy is measured as a function of the incident position, we get a projected map of the muon energy loss.  If this projection is taken from different directions, tomographic reconstruction algorithms, such as filtered back-projection\cite{ct-principle}, can be used to reconstruct the internal structure of the object. Since the reconstruction algorithms are well-studied and beyond our scope, we will demonstrate the result without focusing on the reconstruction.\\
It is worth pointing out that similar approach exists for protons\cite{pct}, but nucleon-nucleon interaction will limit the range of protons for the energy and material of interest, whereas muons interact mainly electromagnetically and are free from such constraint. 
 
\section{Method}
Following the conventions of CT, we use a modulation transfer function (MTF)\cite{spatial-resolution-in-ct} to characterize the performance of muon imaging. MTF measures how the details of different length scales are modulated.  MTF is defined as the Fourier transform of the line-spread function\cite{mtf-intro}, which describes how an infinitely sharp line will be distorted by the imaging system.  The line-spread function is the derivative of the edge-spread function, which describes how an infinitely sharp edge gets distorted. We will start with the edge-spread function.\\
\begin{figure}
	\centering
	\includegraphics[width=0.6\linewidth]{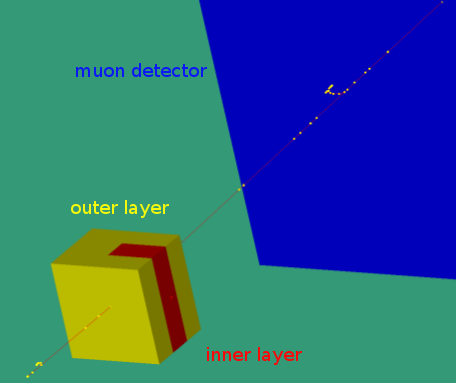}
	\caption{The test object consists of an inner layer sandwiched between two outer layers. Half of the inner layer is made of the same material as the outer layer. Muon beams are scanned through the test object and the energy loss is measured by muon spectrometers.}
	\label{geometry_wtext}
\end{figure}
The test object is two rectangular layers of heavy metal (copper or lead) sandwiching a third layer, half of which is made of the same heavy metal, and the other half-filled with air (see FIG.~\ref{geometry_wtext}). On average muons incident on the heavy metal side lose more energy than the less dense side, and an edge is formed. Normally a numerical derivative should be taken to obtain the line-spread function\cite{mtf-intro}, but we found that the derivative will introduce a large error. So instead we fit the shape of the edge with an error function and the line-spread function is obtained by taking the derivative of the fitted error function. This approach is less prone to numerical errors while retaining the main features. In addition, the Fourier transform of the Gaussian line-spread function is a Gaussian MTF. In the remaining step, the standard deviation of the Gaussian MTF is used as a figure of merit for the imaging resolution and will be referred to as the modulation transfer coefficient (MTC).\\
In the Geant4 simulation\cite{geant4}, $6\times 10^6$ muons are passed through the test object. Muons with the same source position (bin) are averaged and the average muon energy after transmission is plotted as function of source position (FIG.~\ref{example image}).

\section{Result}
\begin{figure}
	\centering
	\begin{subfigure}[b]{0.4\textwidth}
		\includegraphics[width=\textwidth]{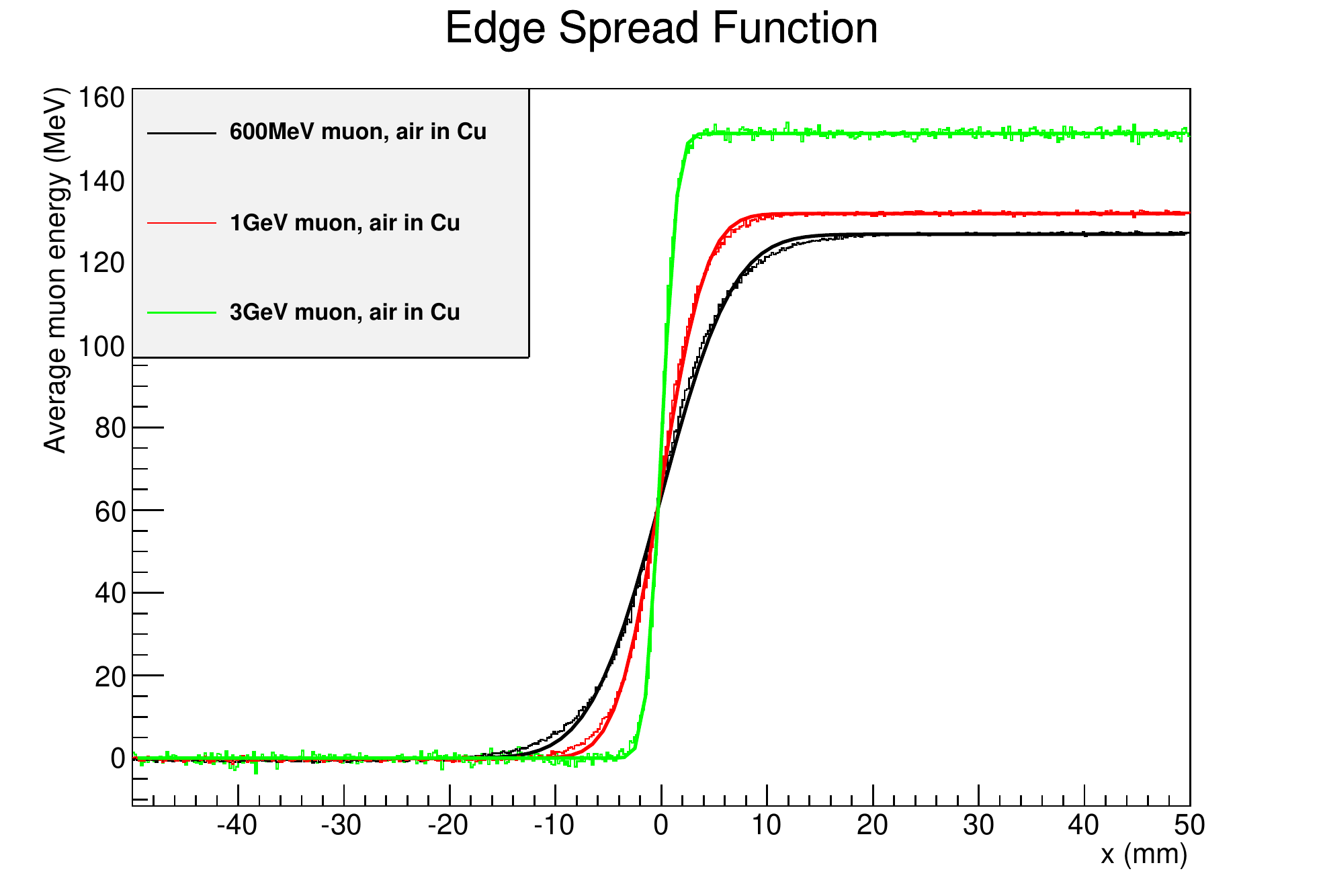}
		\caption{Edge-spread functions.}
		\label{example sub-esf}
	\end{subfigure}
	\begin{subfigure}[b]{0.4\textwidth}
		\includegraphics[width=\textwidth]{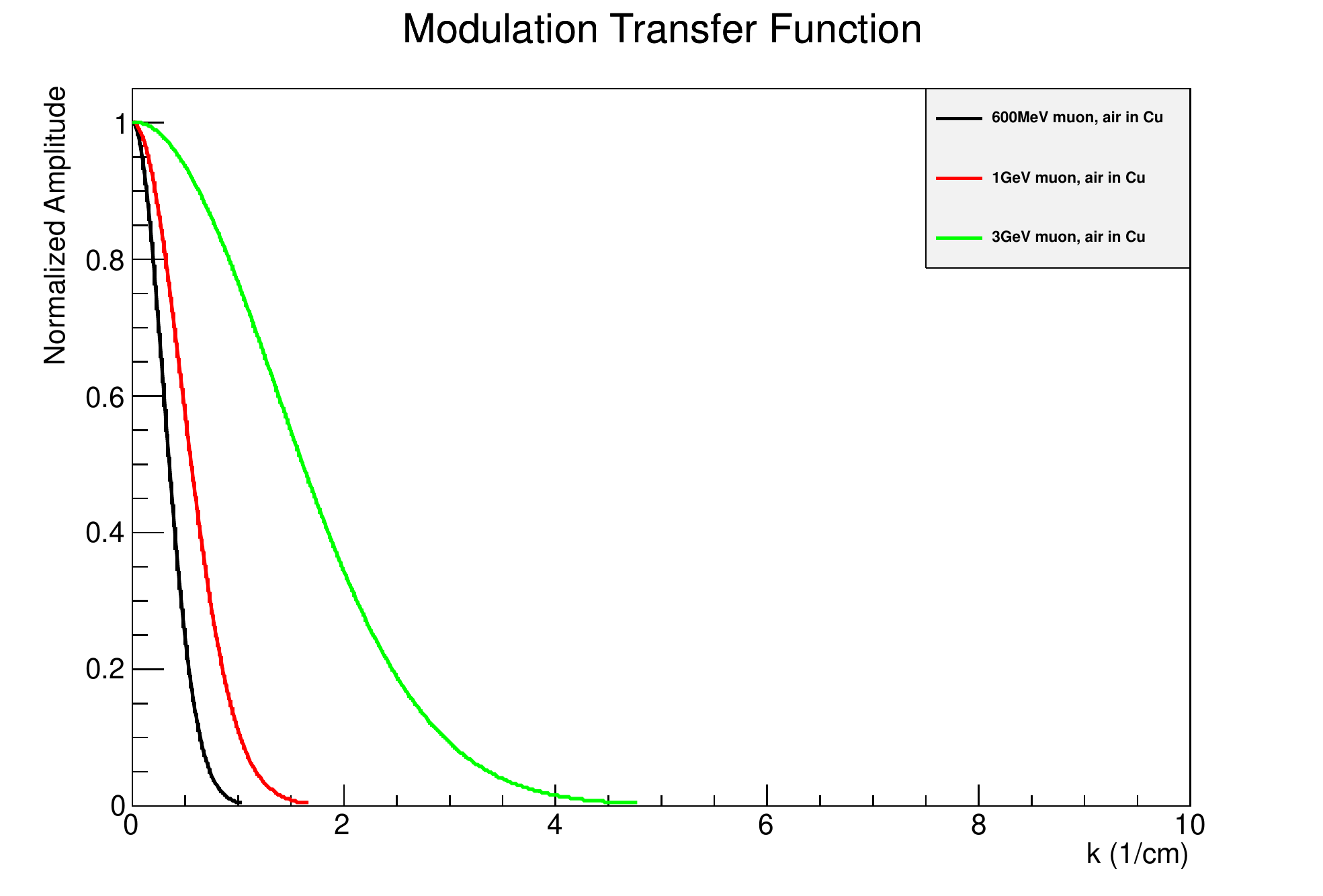}
		\caption{Modulation transfer functions.}
		\label{example sub-mtf}
	\end{subfigure}
	\caption{Edge-spread functions (\ref{example sub-esf}) and MTFs (\ref{example sub-mtf}) for muon energies of $600$~MeV, $1$~GeV and $3$~GeV.  The test object material is copper with an air gap in the middle.  The inner and outer layers are $10$~cm thick.}
	\label{example mtf}
\end{figure}

\begin{figure}
	\centering
	\begin{subfigure}{0.3\textwidth}
		\includegraphics[width=\textwidth]{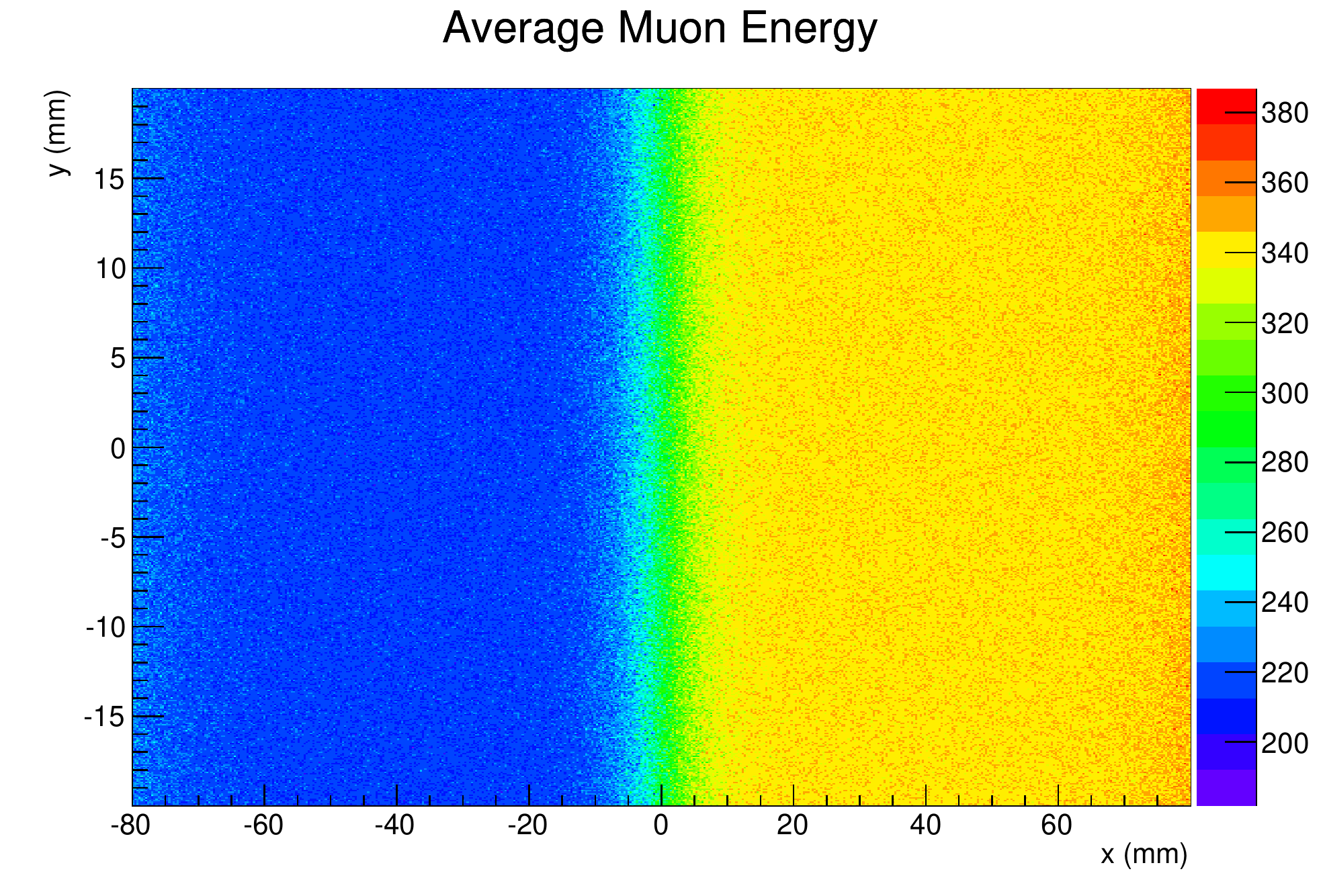}
		\caption{$600$~MeV muons.}
		\label{}
	\end{subfigure}
	\begin{subfigure}{0.3\textwidth}
		\includegraphics[width=\textwidth]{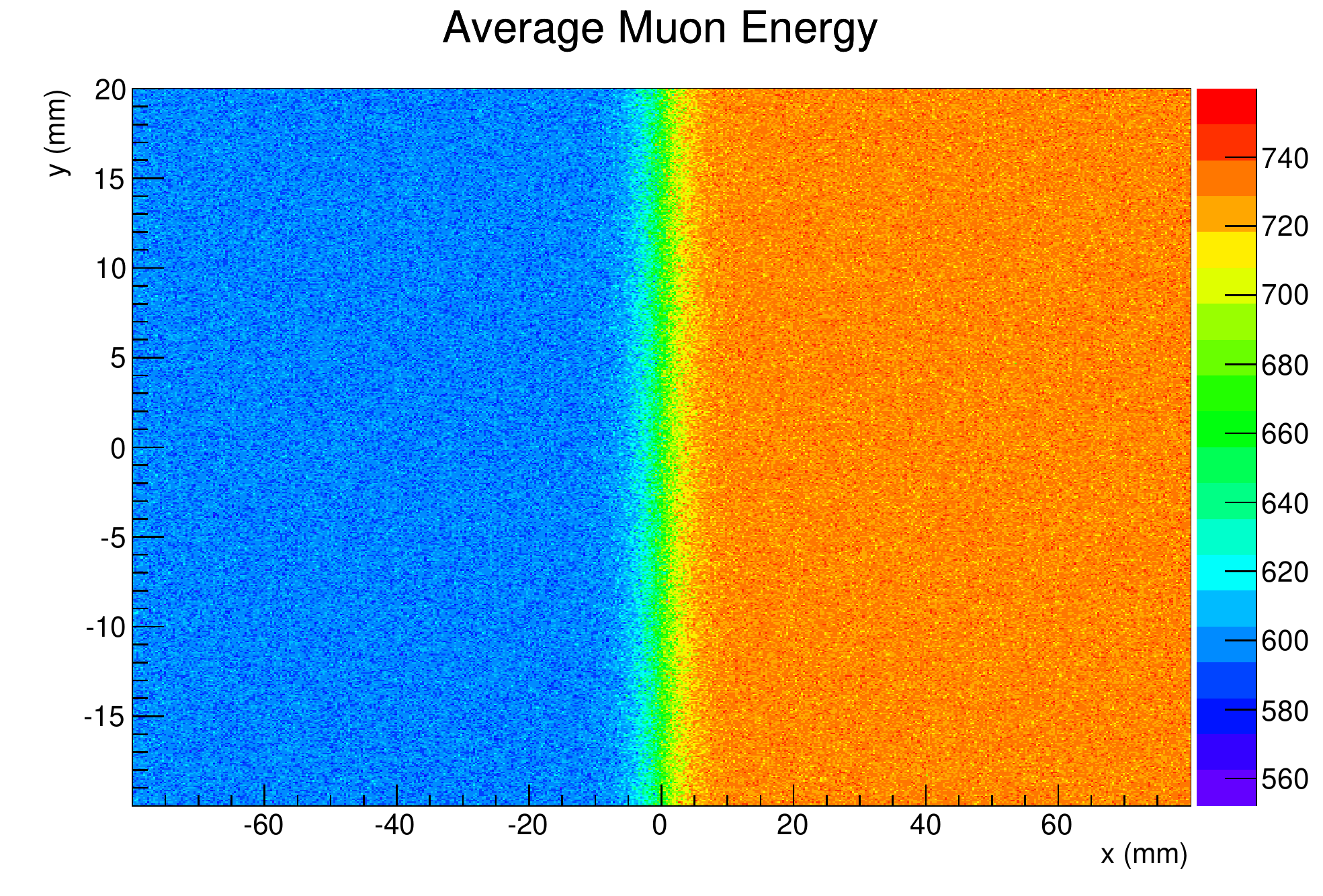}
		\caption{$1$~GeV muons.}
		\label{}
	\end{subfigure}
	\begin{subfigure}{0.3\textwidth}
		\includegraphics[width=\textwidth]{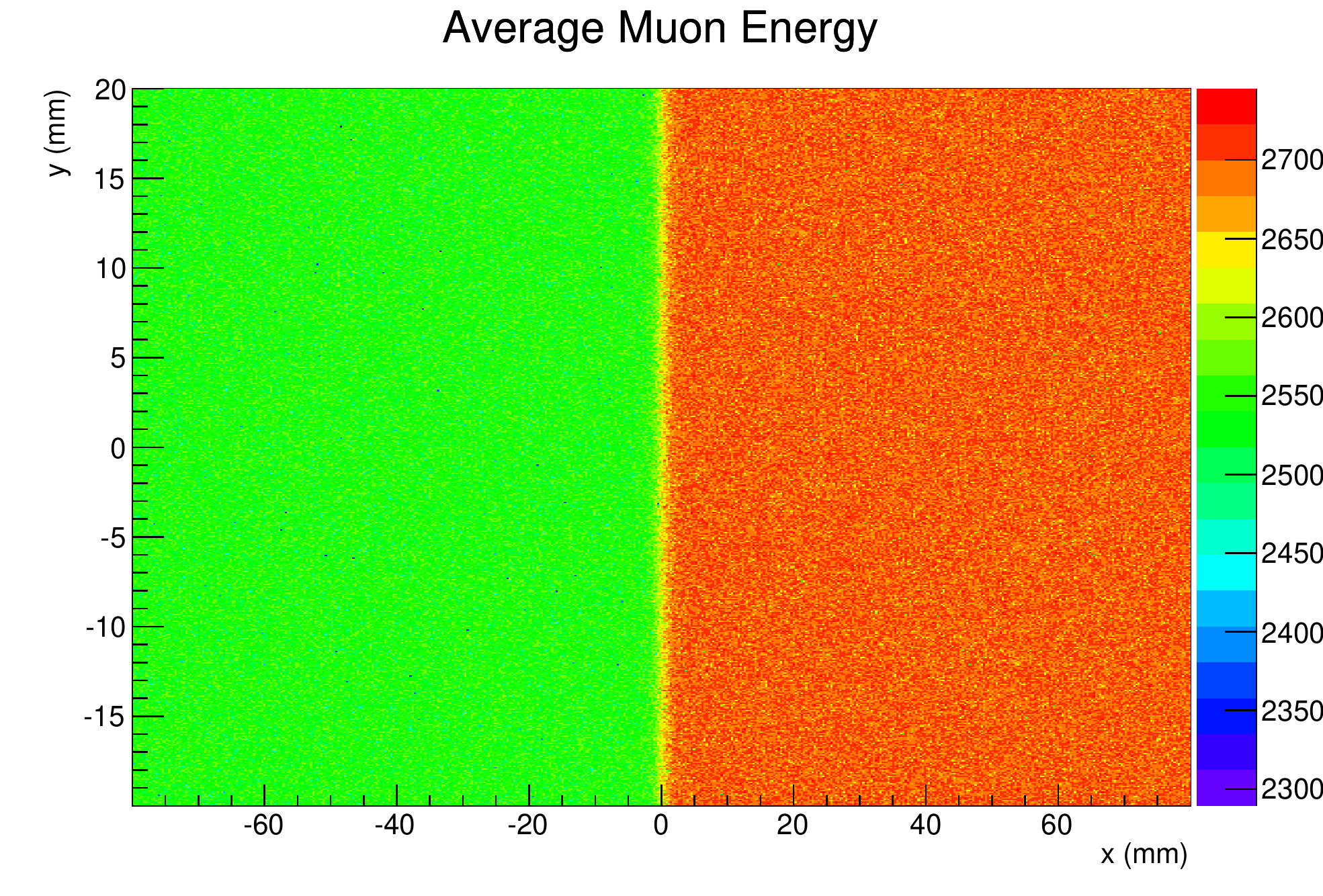}
		\caption{$3$~GeV muons.}
		\label{}
	\end{subfigure}
	\caption{Muon images with different energies incident on a test object with a $10$~cm inner layer of air on the right-half and copper on the left-half and sandwiched between two outer layers of $10$~cm copper plates.  As the incident muon energy is increased, the boundary becomes sharper at the cost of reduced contrast.}
	\label{example image}
\end{figure}

An example edge-spread function and MTF are shown in FIG.~\ref{example mtf}. The corresponding projected images are shown in FIG.~\ref{example image}. From FIG.~\ref{example sub-esf} as the energy is increased, the edge becomes sharper. This is visible in FIG.~\ref{example image}.\\

\begin{figure}
\centering
\includegraphics[width=0.8\linewidth]{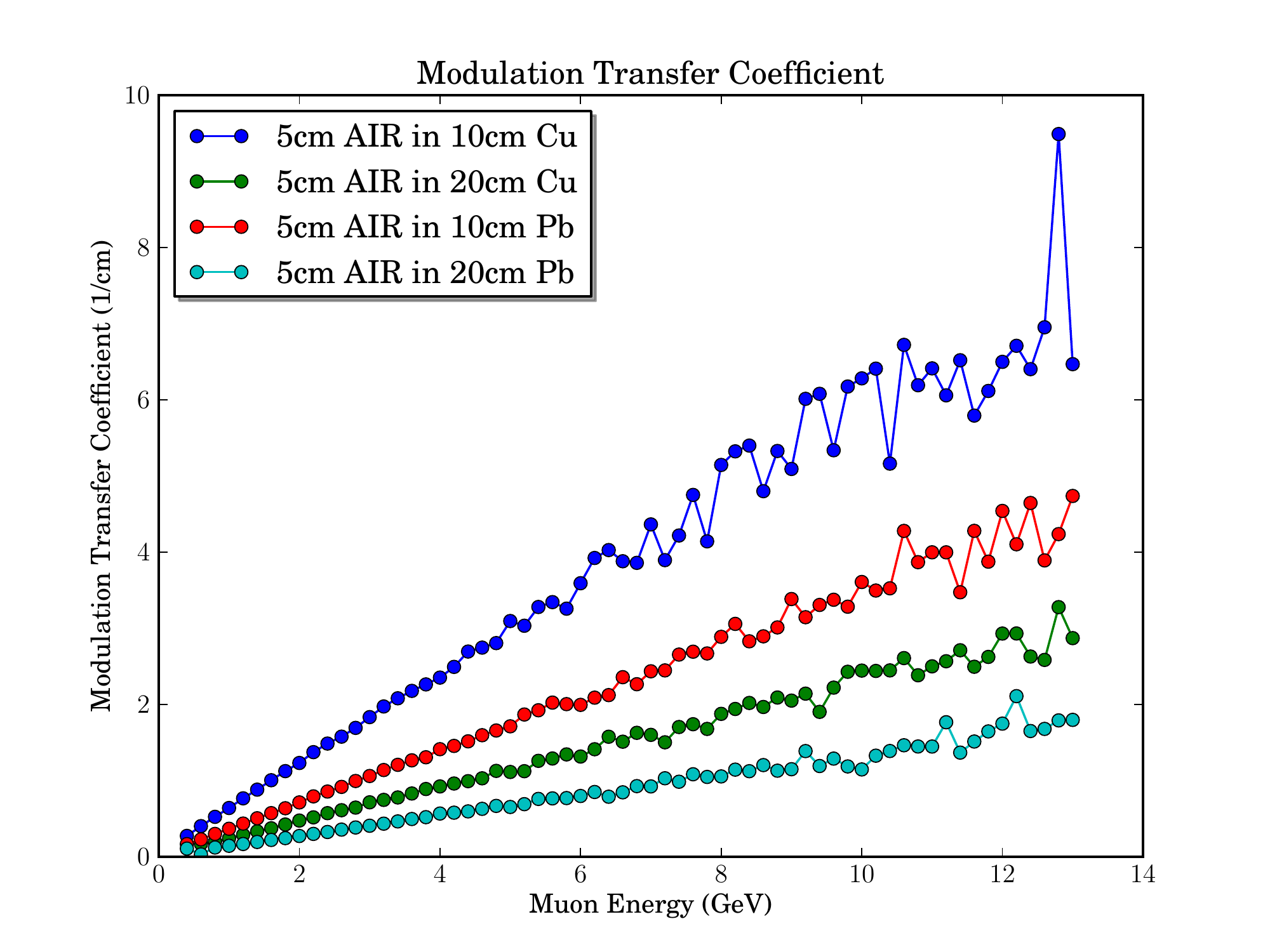}
\caption{The modulation transfer coefficients(MTC) are plotted as a function of muon energy for several test object configurations.  The test object has a $5$~cm inner layer made of copper (or lead) and air. The two outer layers are made of copper (or lead) and each layer is $10$~cm (or $20$~cm) thick.  The MTC increases linearly with incident muon energy and larger statistics are needed at higher energies.  At the same energy, thin layers are better than thick layers, and for the same thickness copper is better than lead.}
\label{mtf}
\end{figure}

\begin{figure}
\centering
\includegraphics[width=0.8\linewidth]{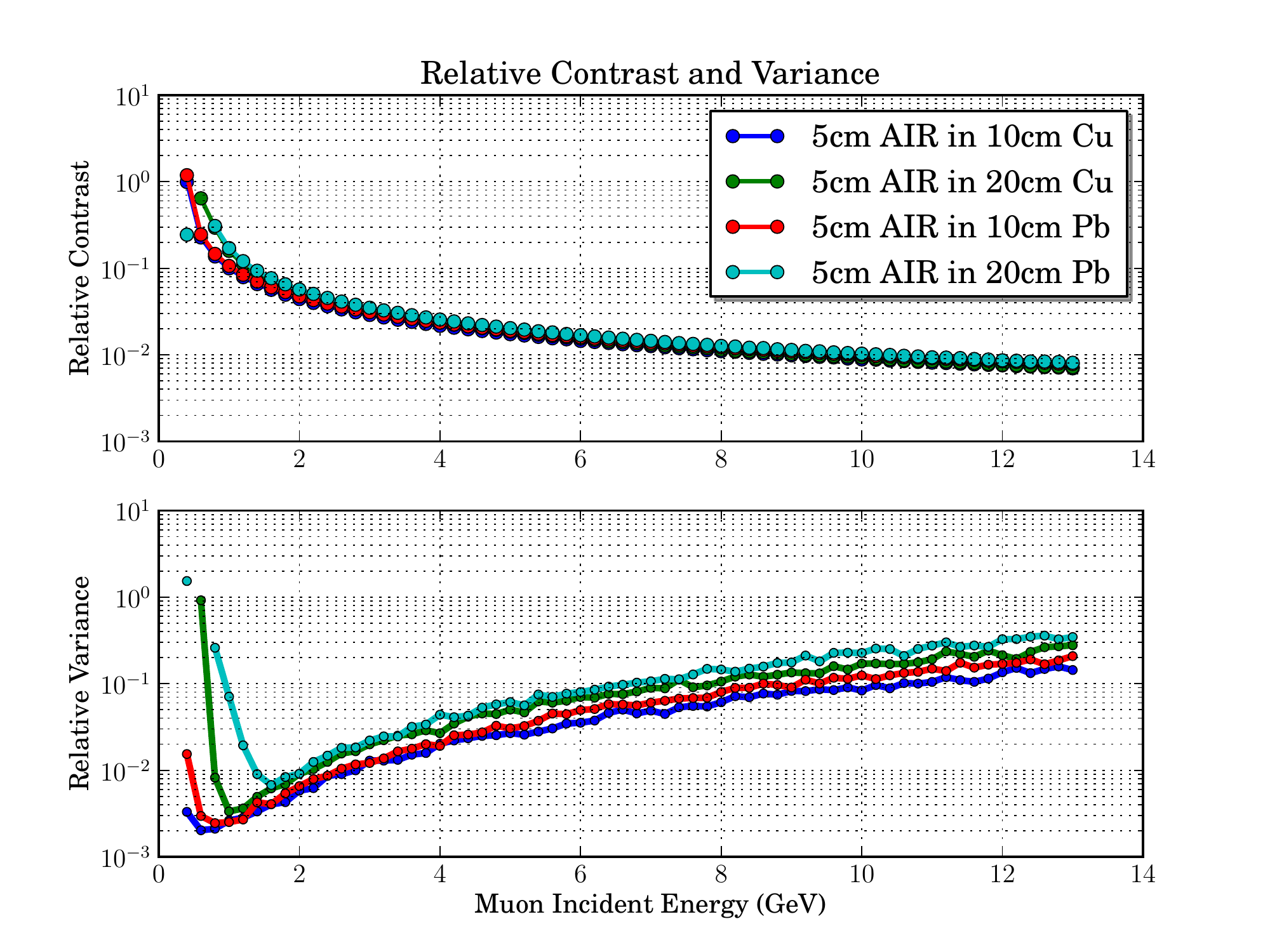}
\caption{Relative amplitude (top) and relative variance (bottom) as function of incident muon energy. As the muon energy increases, the relative amplitude decreases while the relative variance first reaches a minimum and then increases. Low relative amplitude requires higher detector resolution, and large variance requires more statistics.}
\label{relative amplitude}
\end{figure}

The MTC as a function of the incoming muon energy is shown in FIG.~\ref{mtf}.   The resolution increases with muon energy, and thin layers are better than thick layers. For the same thickness and muon energy, copper is better than lead. These findings are consistent with the expected effects of multiple scattering.
The broadening is inversely proportional to the momentum of the muon and 
increases with the square root of the thickness of the material in units of radiation length\cite{bethe-bloch}.\\

The relative amplitude, defined as the ratio of the amplitude of the fitted error function to the vertical offset, is shown as function of muon energy in FIG.~\ref{relative amplitude}. As the muon energy increases, the relative amplitude (contrast) decreases. The relative variance, defined as the variance with respect to the fitted error function divided by the amplitude increases after reaching a minimum.   For high energy muons, image resolution is improved at the cost of requiring better detector resolution to resolve the reduced contrast.  Larger statistics are required to reduce fluctuations in a high resolution image.\\

\begin{figure}
\centering
\includegraphics[width=0.8\linewidth]{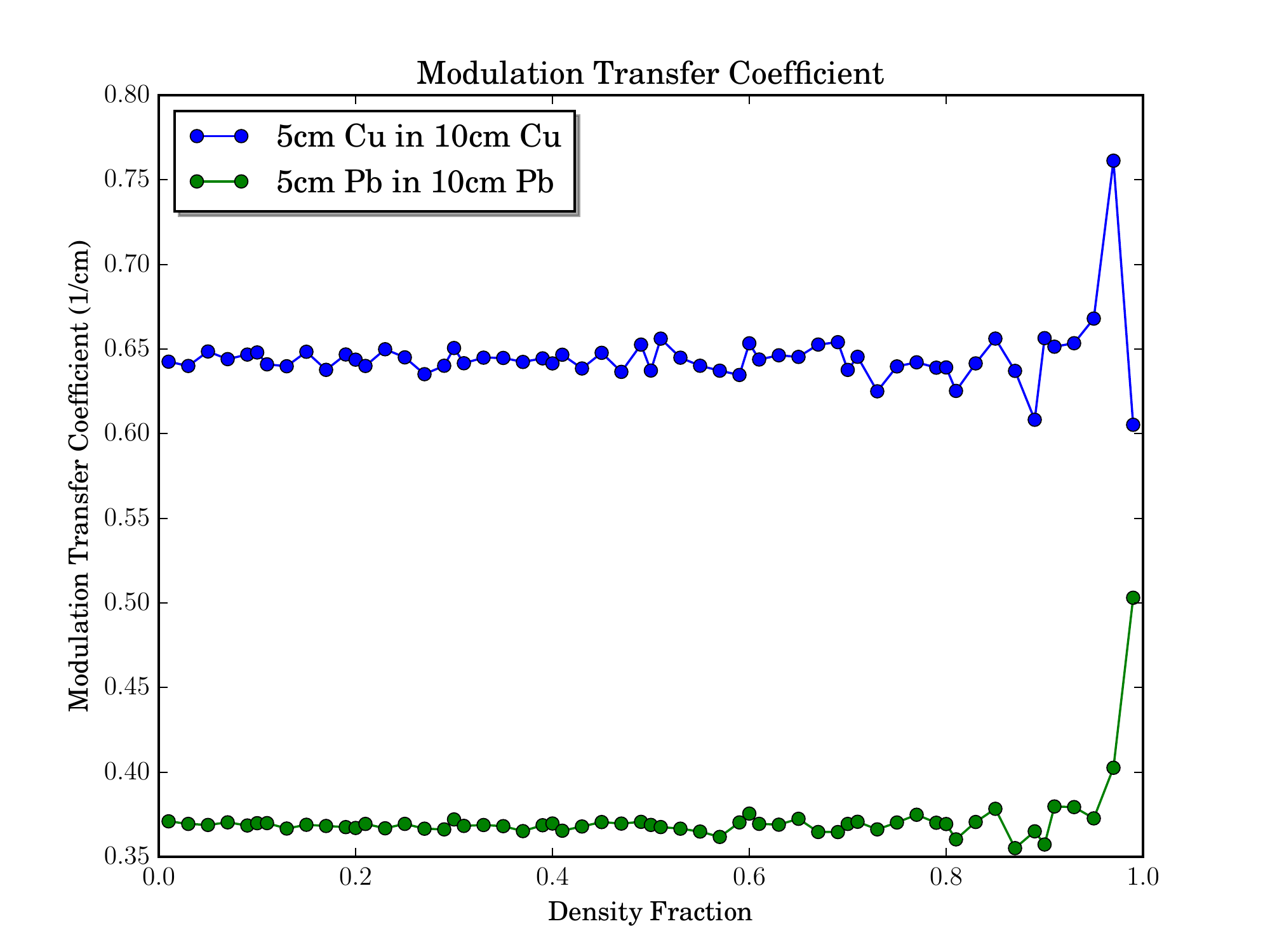}
\caption{The top (bottom) curve shows MTC for $5$ cm copper (lead) of different densities with $10$ cm copper (lead) layers. The muons have energy $1$ GeV. The density fraction refers to the scale to the normal density of respective materials. At a density fraction close to $1$, the fluctuation grows since it becomes harder to distinguish the two different densities. While copper and lead have quite different MTCs, they stay constant for a large range of densities.}
\label{mtf different density}
\end{figure}

\begin{figure}
\centering
\includegraphics[width=0.8\linewidth]{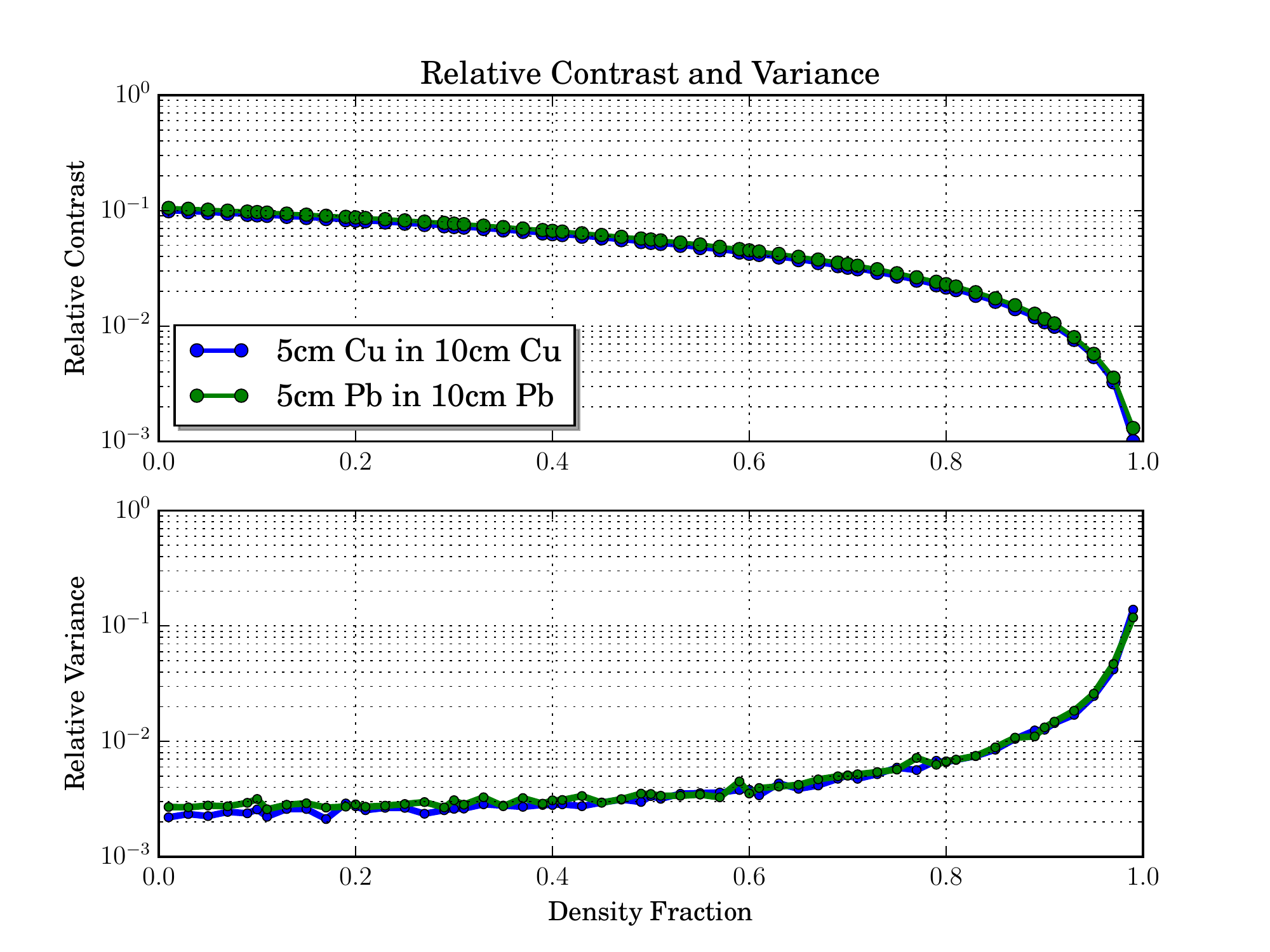}
\caption{Contrast and variance as function of density fraction. As density fraction gets close to 1, the contrast drops and the variance increases. Since the MTC stays approximately constant, the ability to distinguish materials of the same type but different density is limited by the detector resolution.}
\label{amplitude different density}
\end{figure}

Sensitivity to density transition edges were tested for $1$ GeV muons, a $5$~cm inner layer, $10$~cm outer layers, and where the air gap is replaced with copper (lead) of different densities. The density is scaled with respect to the normal density of copper (lead). The MTC as a function of density shown in FIG.~\ref{mtf different density}.  The MTCs are largely independent of the densities, except when the densities are the same to within 5\%.   The relative amplitude as a function of density fraction is shown in FIG.~\ref{amplitude different density}.  When the imaging materials are of the same type but have different densities, the closer the densities are, the higher the required detector resolution.  For example, if we had a muon detector with a resolution of $2$\% at $1$~GeV, we can resolve $8.9$ g/cm$^3$ copper from copper scaled to a density of $7.12$ g/cm$^3$.

\begin{figure}
\centering
\includegraphics[width=\linewidth]{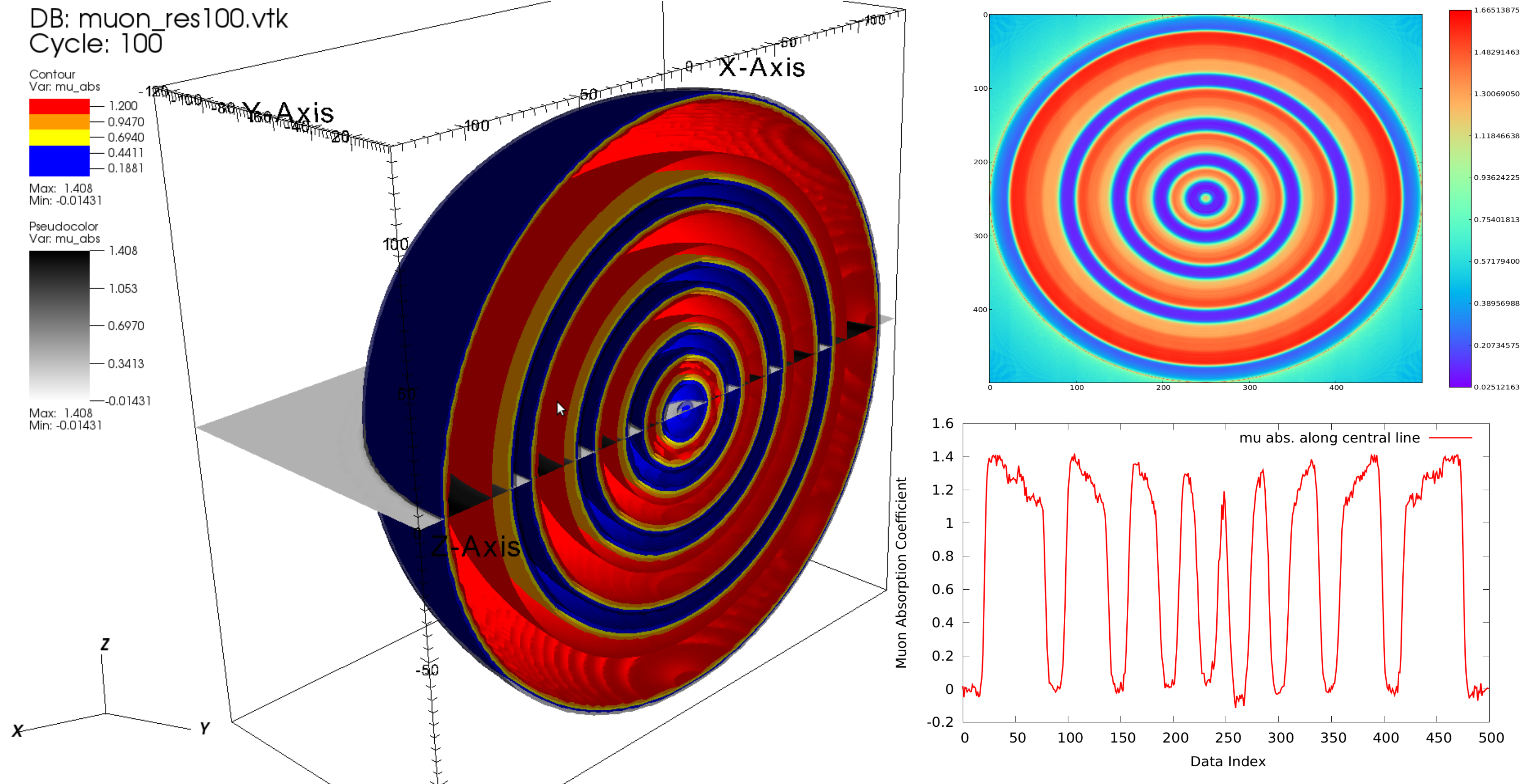}
\caption{(left) 3D reconstruction of a test object using VisIt\cite{visit}, taken with $3$ GeV muons. The object consists of $5$ concentric spheres, with thickness $3$, $9$, $15$, $21$, and $30$~mm, respectively, and $10$~mm of water between different metallic layers. Each metallic layer consists of lead, copper and iron of equal thickness. (right) Cross-section view of the test object, and the muon absorption along a radial line. For the outer layers, we can identify three distinct materials indicated by the color in the cross-section, and the steps in the line plot.}
\label{demo}
\end{figure}

In FIG.~\ref{demo} we demonstrate a $3$D reconstruction of a test object made of concentric spheres of lead, iron and copper, together with its cross-section image and radial distribution of muon absorption coefficient. Between different layers there is $1$~cm of water. Reconstruction was done with filtered back-projection\cite{ct-principle} with $200$ angles.

\section{Discussion}

The resolution is limited by the multiple scattering of muons inside the materials, so any measure that reduces muon scattering will improve the image resolution. Once the geometry is fixed, the only available option is to increase the muon energy\cite{bethe-bloch}. However, this has a cost of reduced contrast and increased noise, and requires detectors with higher resolutions and a larger statistics of muons. 
This trade-off can benefit from a spectrum of muon momenta.  One can scan the material with lower energy muons to determine the approximate regions and then switch to higher energy to determine the boundaries with better resolution. In these studies, no selection is made. Resolution can be further improved by discarding muons with large amount of scattering.\\

Since muon tomography records the information of individual muons, one can use the amount of muon scattering to identify internal boundaries where variance come to local maximum, and to identify different materials. The latter has been demonstrated with cosmic muons\cite{material-id} and can be applied to accelerator muons.  In addition, for muons that are monochromatic or have a well-measured incident momentum, from the measured energy loss one can use he Bethe-Bloch equation (Eq.~\ref{bethe-block}) to identify materials or test models of different material compositions. In this way, one can examine the radiography of an object.\\

Muon imaging systems can be co-located at neutrino beam facilities. In accelerator-based neutrino experiments, muon neutrinos are obtained from charged pion decays along with muons\cite{numi-design}. However, these muons are typically filtered from the beam without being used. Given the accelerator neutrino energy, the muon energy can be calculated. Let $\theta$ denote the angle between neutrino and the pion and assuming zero neutrino mass, neutrino energy is:
	\begin{eqnarray}
		E_\nu &=& \frac{m_\pi^2-m_\mu^2}{2E_\pi-2p_\pi \cos\theta}  \label{v energy}\\
		&\approx& \frac{(1-m_\mu^2/m_\pi^2)}{1+\gamma^2\theta^2}E_\pi. \text{ (small angle)} \label{v approx}
	\end{eqnarray}	 
A similar calculation is done in \cite{pion_decay_kinematics}. In the center-of-mass frame, the neutrino angular distribution is isotropic, so in the lab frame the neutrino direction has angular distribution 
$P(\theta)=\sin\theta/[\gamma^2(1-\beta\cos\theta)^2] \label{angular distribution}$, 
with the most likely angle for neutrino $\theta^*=\cos^{-1}\beta$.
The ratio of muon energy to the neutrino energy for the most probable angle is:
	\begin{eqnarray}
		E_\mu/E_\nu &=& \frac{2m_\pi^2}{m_\pi^2-m_\mu^2}-1\approx 3.7. \label{ratio}  
	\end{eqnarray}
Taking NuMI at Fermilab as an example, the neutrino beam energy is in the range $1-3$~GeV for the low energy option\cite{minos-flux,arxiv-muon-flux}. The corresponding muon energy is $4-10$~GeV, which lies in the energy range of a muon computed tomography system. The CNGS at CERN has about $3$ times larger energy\cite{cngs-cern,cngs-icarus}. The required muon energy is different for each object and material, but $10$ GeV is enough for imaging most objects of moderate size. If there is a need to reduce the muon energies coming out of the accelerator facility, then the muon beam can be cooled down with an absorber and selected to the desired energy with a spectrometer. \\

The muon flux produced in associated with the production of neutrino beams at the NuMI facility
is roughly $10^7$/cm$^2$ per spill, over a roughly $1$~m$^2$ beam area\cite{numi-design,minos-flux}.  The NuMI spill cycle is a spill duration of approximately $10$~$\mu$s every $1.87$ s\cite{minos-flux}.  In between each spill, the orientation of the object with respect to the
beam can be stepped in angular increments.  For one degree steps in azimuthal and polar
directions, the imaging data for a roughly $1$~m$^2$ cross-sectional area can be collected in 
roughly $12$ minutes. The test object reconstruction in FIG.~\ref{demo} assumes approximately $20$~min beam time. The limitation of such a device is largely instrumental.  To simultaneously
acquire such a large flux of muon data requires highly pixelated silicon tracking operating at 
hundreds of MHz.  There are silicon pixel readout systems with these capabilities designed for high-rate environments\cite{chistiansen2013rd}. 
Spectrometer selection could aid in reducing the flux by selecting a narrow
band of incident muon momenta.

\section{Conclusion}
In this paper, we have shown that computed tomography with accelerator muons can be used in place 
of X-rays to overcome the limitations of imaging inside of objects made of heavy metals such as copper and lead. The spatial resolution is shown to increase with increasing muon energy, due to the reduction in 
multiple scattering. However, this reduction is at the cost of image contrast and requires more statistics to suppress fluctuations. The resolution also depends on the geometry, material and material boundaries. In the case of copper and lead, lead scatters muon more, and therefore results in poorer resolution. For materials of the same type but different densities, the resolution remains somewhat constant as a function of the density fraction. 
The relative amplitude decreases as the densities get closer, and, therefore, places a more stringent requirement on the detector resolution to distinguish regions having different densities. Either by measuring the multiple scattering or energy loss of individual muons, it is possible to do material identification.\\

We have framed the capabilities of muon computed tomography through simulation.  As with X-ray CT
at the outset, it takes foresight and planning to envision the breakthroughs that dense object imaging may yield.  Muon CT is a probe like no other into the unknown that hides deep within dense structures.
The muons produced in association with high intensity neutrino beams fall into the energy range of interest for muon imaging.  This provides a unique opportunity to incorporate muon imaging systems into existing or future neutrino beam facilities.

\begin{acknowledgments}
This project was supported under DOE Award No. ER-41850 and a Princeton University Dean of Research Innovation Fund for Research Collaborations between Artists and Scientists Award (2014). The authors are pleased to acknowledge that the work was substantially performed at the TIGRESS high performance computer center at Princeton University which is jointly supported by the Princeton Institute for Computational Science and Engineering and the Princeton University Office of Information Technology's Research Computing department.
\end{acknowledgments}
\bibliographystyle{apsrev}
\bibliography{muon}

\end{document}